\documentclass[aps,prl,twocolumn,amsmath,amsfonts,showpacs]{revtex4}

\usepackage{graphicx}
\usepackage{bm}
\newcommand{\nn}{\nonumber}
\newcommand{\beq}{\begin{eqnarray}}
\newcommand{\eeq}{\end{eqnarray}}
\newcommand{\bpm}{\begin{pmatrix}}
\newcommand{\epm}{\end{pmatrix}}

\newcommand{\p}{\partial}
\newcommand{\e}{\epsilon}
\newcommand{\s}{\sigma}

\newcommand{\D}{\Delta}
\begin{document}

\title{
Negative thermal magnetoresistivity as a signature of chiral anomaly in Weyl superconductors

}


\author{Takuro Kobayashi}
\author{Taiki Matsushita}
\author{Takeshi Mizushima}
\author{Atsushi Tsuruta}
\author{Satoshi Fujimoto}
\affiliation{Department of Materials Engineering Science, Osaka University, Toyonaka 560-8531, Japan}




\date{\today}

\begin{abstract}
We propose that 
chiral anomaly of Weyl superconductors gives rise to negative thermal magnetoresistivity
induced by emergent magnetic fields, which are generated by vortex textures of order parameters or lattice strain.
We establish this scenario by combining the argument based on Berry curvatures,
and the quasi-classical theory of the Eilenberger equation with quantum corrections arising from inhomogeneous
structures. It is found that the chiral anomaly contribution of the thermal conductivity exhibits characteristic temperature dependence, which can be a smoking-gun signature of this effect.
\end{abstract}



\maketitle


{\it Introduction.}--- In Weyl semi-metals and Weyl superconductors, low-energy excitations behave as Weyl fermions characterized by nonzero Berry curvatures in the momentum space, which stem from monopole charges at Weyl points \cite{vishwanath,balents,balents2,murakami,volovik,volovik2,volovik3,Hosur_review,PhysRevB.86.054504}.
This feature results in various intriguing electromagnetic responses associated with chiral anomaly. For instance, in the case of Weyl semi-metals, chiral anomaly gives rise to the anomalous Hall effect, chiral magnetic effect, and negative magnetoresistivity  \cite{NN_1983, Fukushima_CME2008,Vazifeh2013,Zyuzin_AHE2012,Liu2013,PhysRevB.88.104412, Goswami2013,Lucas9463}, some of which
have been already experimentally verified in real materials \cite{volovik4,Xu613,LvTaAs2015,TaAsWeng2015,huang-nat,XiaochunTaAs2015,Shekhar2015}.
For Weyl superconductors, however, chiral anomaly phenomena can not be realized by simply applying electromagnetic fields, because
Weyl-Bogoliubov quasiparticles do not carry definite charges. Instead, chiral anomaly in the superconducting state can be induced by
emergent electromagnetic fields which are generated by spatially inhomogeneous textures of order parameters, or lattice strain \cite{guinea,PhysRevB.92.165131,T.Hughes2011,T.Hughes2013,Parrikar2014,Chandia,Shitade2014,Gromov2014,PhysRevLett.115.177202,PhysRevLett.116.166601,PhysRevB.94.241405,PhysRevX.6.041046,PhysRevX.6.041021,PhysRevB.96.081110,PhysRevB.95.041201,PhysRevB.96.224518}.

In this letter, we demonstrate that negative magnetoreisistivity of longitudinal thermal currents induced by an emergent magnetic field can be a signature of chiral anomaly; i.e., 
thermal conductivity of Weyl quasiparticles increases as the emergent magnetic field parallel to the temperature gradient increases, even when pair-breaking effects due to magnetic fields are negligibly small.
We examine two scenarios for realizing emergent magnetic fields. One is that induced by vortex textures in the mixed state,
and the other one is a chiral magnetic field arising from lattice strain \cite{PhysRevB.92.165131,PhysRevB.95.041201,PhysRevB.96.224518}. 
We establish the above-mentioned result by combining the argument based on the semiclassical equation of motion with Berry curvatures characterizing Weyl fermions, and microscopic analysis using quasiclassical theory of the Keldysh Green function.
 Our finding is relevant to putative Weyl superconductors such as multi-layer systems \cite{PhysRevB.86.054504}, and uranium-based systems, URu$_2$Si$_2$, UPt$_3$, 
 UCoGe, U$_{1-x}$Th$_{x}$Be$_{13}$~\cite{sato-fujimoto,PhysRevLett.99.116402,doi:10.7566/JPSJ.85.033704,PhysRevB.91.140506,URuSi1,goswami,PhysRevB.92.214504,Schemm190,PhysRevLett.108.066403,PhysRevB.66.134504,shimizu,mizushimaPRB18,machidaJPSJ18}.

{\it Semiclassical argument for thermal transport with Berry curvature.}--- We, first, present a semiclassical argument for thermal transport.  This approach is useful for qualitative understanding of chiral anomaly effects.
We consider a paradigmatic model of Weyl superconductors which describes a three-dimensional (3D) chiral $p_x+ip_y$ pairing state of spinless fermions, 
though our basic idea can be generalized to any Weyl superconductors.
The superconducting gap function for homogeneous cases is given by $\Delta_{\bm{k}}=\Delta(k_x-ik_y)/k_F$.
In this system, low-energy excitations from point nodes of the superconducting gap 
at $\bm{k}=(0,0,\pm k_F)$ behave as Weyl fermions. 
 The model Hamiltonian for low-energy Weyl quasiparticles with the monopole charge $s=\pm 1$
 in the case with spatial inhomogeneity is given by,
 \begin{eqnarray}
\mathcal{H}_{s}(\bm{k},\bm{r})=s e^{\mu}_aV^a_b\tau^b(k_{\mu} - sk_{0\mu}),
\label{eq:ham1}
\end{eqnarray}
where $V^a_b=\mbox{diag}[\frac{\Delta}{k_F},\frac{\Delta}{k_F},v_F]$ with $v_F$ the Fermi velocity,
$\tau^a$ is the Pauli matrix in the particle-hole space.
Spatial inhomogeneity is described in terms of the vielbein $e^{\mu}_a$.
We use greek letter indices $\mu=1,2,3$ as space indices for the laboratory frame, and roman letters $a=\bar{1},\bar{2},\bar{3}$ as indices for a local orthogonal frame. 
As mentioned above,  the spatial inhomogeneity gives rise to an emergent magnetic field $\bm{\mathcal{B}}=\bm{T}^{\mu}k_{\mu}$ with
the torsion field, 
$
(\bm{T}^{\mu})^{\nu}=\frac{1}{2}\epsilon^{\nu\lambda\rho}T^a_{\lambda\rho}e_a^{\mu}
$, 
$
T^a_{\mu\nu}=\partial_{\mu} e^a_{\nu}-\partial_{\nu} e^a_{\mu}
$,
where $e^a_{\mu}$ is the inverse of $e^{\mu}_a$ \cite{PhysRevB.92.165131,T.Hughes2011,T.Hughes2013,Parrikar2014,SM}.
It is noted that $\bm{\mathcal{B}}$ plays a role of a chiral magnetic field, when $\bm{T}^{z}$ is nonzero,
since the sign of $k_z$ at the Weyl points of the model (\ref{eq:ham1}) corresponds to chirality of Weyl fermions.
There are several ways of realizing nonzero $\bm{\mathcal{B}}$ in superconductors.
For instance, 
a vortex line texture parallel to the $z$-axis, i.e. $\Delta=\Delta_0e^{i\phi}$ generates the emergent magnetic field,
$\bm{\mathcal{B}}=(0,0,\mathcal{B}_z)$ with
$
\mathcal{B}_z=T_{12}^{\mu}k_{\mu}=(k_y\cos \phi -k_x \sin\phi )/r,
$
which does not depend on  $k_z$, and is not a chiral magnetic field, but imitates a usual magnetic field.
Also, lattice strain such as twist of a crystal structure with a rotation axis parallel to $z$-direction gives rise to an emergent chiral magnetic field along the $z$-axis. 
In the following, we consider magnetoresistivity of thermal current for these two cases. 

By using the semiclassical equation of motion with Berry curvatures for Weyl quasiparticles~\cite{SM}, and the Boltzmann equation,
we obtain 
 the chiral anomaly contribution of the local thermal current $\bm{J}_H(\bm{r})$  up to 
 leading terms in $\bm{\mathcal{B}}$,
\begin{eqnarray}
\bm{J}_H(\bm{r})
&=&\sum_{s=\pm 1}\sum_{\bm{k}}(\bm{v}_{\bm{p}s}\cdot\bm{\Omega}_{\bm{kk}s})^2\varepsilon_{\bm{k}s}^2\left(\frac{\partial f}{\partial \varepsilon_{\bm{k}s}}\right)\tau_{\bm{k}s} \nonumber \\
&&\times\left(\frac{\nabla T}{T}\cdot\bm{\mathcal{B}}\right)\bm{\mathcal{B}},
\label{eq:hc}
\end{eqnarray}
where $\varepsilon_{\bm{k}s}=\sqrt{v^2(k_z-sk_{0z})^2+\Delta^2(k_x^2+k_y^2)/k_F^2}$, $\bm{v}_{\bm{k}s}=\partial \varepsilon_{\bm{k}s}/\partial\bm{k}$, 
$\tau_{\bm{k}s}$ is the relaxation time, $f$ is the Fermi distribution function, and $\bm{\Omega}_{\bm{k}\bm{k}s}$
is the Berry curvature generated by the monopole charge at the Weyl point,
 which characterizes the chiral anomaly contribution.
Equation (\ref{eq:hc}) evidences the negative thermal magnetoresistivity  (NTMR) 
due to the emergent magnetic field $\bm{\mathcal{B}}$.
It is noted that the chiral anomaly contribution of the thermal conductivity $\kappa_A$ extracted from Eq.~(\ref{eq:hc})
 exhibits singular temperature dependence. In the case of a constant relaxation time, we have,
 \begin{eqnarray}
  \kappa_A \propto 1/T,
  \label{eq:tc2}
  \end{eqnarray} 
  for low $T$.
  If one takes into account temperature-dependence of $\tau_{\bm{k}s}$ more precisely,
  the low-temperature behavior becomes more singular.
 This behavior is due to the singularity of the Berry curvature
 in the vicinity of Weyl points, i.e. $\Omega_{\bm{kk}s} \sim 1/|\delta\bm{k}|^2$ for the deviation from the Weyl points $|\delta\bm{k}|\rightarrow 0$.
The characteristic $T$-dependence of (\ref{eq:tc2}) can be utilized for discriminating the chiral anomaly contribution from usual contributions of thermal conductivity of nodal excitations, $\kappa_0 \propto T$ for $T\rightarrow 0$.
However, we must be careful about the applicability of Eq.~(\ref{eq:hc}). 
The divergent behavior of (\ref{eq:tc2}) implies that it can not be used in the low-temperature limit, for which adiabatic approximation postulated for the derivation of the Berry curvature formula fails down.
Thus, Eq.~(\ref{eq:tc2}) is applicable only in the intermediate temperature region. 
To investigate thermal transport for the whole temperature region, we exploit alternative approaches based on 
the Keldysh formalism
in the following.

{\it Keldysh-Eilenberger approach for cases with vortex textures.}--- To confirm the prediction obtained above, and go beyond adiabatic approximation, which fails down in the low-temperature region, we exploit the Keldysh 
formalism of the quasiclassical Eilenberger equation.
We consider the 3D chiral $p_x+ip_y$ pairing model again, and, first, examine the case of an emergent magnetic field generated by
vortex textures of the superconducting order parameter. The case of strain-induced chiral magnetic fields will be considered later.
A merit of the scenario of a vortex-induced emergent magnetic field is that it can be easily realized for any type-II superconductors.
Transport properties of systems with inhomogeneous textures are described in terms of the quasiclassical 
Green function  $\check{g}(\hat{\boldsymbol{k}},\boldsymbol{r},\epsilon)$  
with $\hat{\bm{k}}$ a unit vector parallel to the Fermi momentum.\cite{SM,rainer,graf,eschrig}.
Using the Keldysh Green function $\hat{g}^K$, we can express
a thermal current as,
\begin{equation}
\boldsymbol{J}_H(\boldsymbol{r})=N_F\int_{-\infty}^{\infty}{\frac{d\epsilon}{4\pi i}}
\int{d\hat{\boldsymbol{k}}\epsilon {\bm v}_{F}\frac{1}{2}\mathrm{Tr}\left[\hat{g}^K(\hat{\boldsymbol{k}},\boldsymbol{r},\epsilon)\right]},
\label{eq:hc2}
\end{equation}
where $N_F$ is the density of states at the Fermi level, ${\bm v}_{\rm F}$ is the Fermi velocity, and $\int d\hat{\bm k}\cdots$ is the normalized Fermi surface average. In this paper, we consider the spherical Fermi surface with ${\bm v}_{\rm F}=v_{\rm F}\hat{\bm k}$.

Effects of emergent magnetic fields arising from spatial inhomogeneity can be incorporated via spatial gradient expansion
of the Eilenberger equation, which gives
higher-order quantum corrections to the quasiclassical approximation. 
Up to the first order in $1/(k_F\xi)$ with $\xi$ the coherence length,
the Eilenberger equation with quantum corrections 
is given by~\cite{SM},
\begin{align}
[(\epsilon+e\bm{v}_F\cdot\bm{A})\tau_3-\check{h},\check{g}]+i\boldsymbol{v}_F\cdot\boldsymbol{\nabla}_{\boldsymbol{r}}\check{g}=\frac{i}{2}\{\check{h}\cdot\check{g}\}-\frac{i}{2}\{\check{g}\cdot\check{h}\}, 
\label{eq:ee1}
\end{align}
where  
$
\{\check{a}\cdot\check{b}\}=\boldsymbol{\nabla}_{\boldsymbol{r}}\check{a}\cdot\boldsymbol{\nabla}_{\boldsymbol{k}}\check{b}-\boldsymbol{\nabla}_{\boldsymbol{k}}\check{a}\cdot\boldsymbol{\nabla}_{\boldsymbol{r}}\check{b}
$,
and $\bm{A}$ is a vector potential due to an external magnetic field, 
and $\check{h}=\check{\Delta}+\check{\sigma}_{\rm imp}$ with $\check{\Delta}$ the gap function, and  $\check{\sigma}_{\rm imp}$ the self-energy due
to impurity scattering, which determines the relaxation time $\tau$ \cite{SM}. 
The nonzero right-hand side term of (\ref{eq:ee1}) describes leading quantum corrections.
For simplicity, we assume that  $\check{\sigma}_{\rm imp}$ does not depend on temperature $T$.
In general,  $\check{\sigma}_{\rm imp}$ should depends on $T$, 
because of the energy-dependence of the density of states
of Weyl quasiparticles, and $T$-dependence of the gap function. 
However, this simplification is useful for the investigation of characteristic 
 $T$-dependence of thermal conductivity arising from chiral anomaly, which is predicted from the semiclassical analysis (\ref{eq:tc2}).
 Effects of an emergent magnetic field caused by vortex textures are included in the right-hand side of Eq.~(\ref{eq:ee1}).
We deal with this term in a perturbative way. 
We expand the Green function up to the second order in $1/(k_F\xi)$;
$\check{g}=\check{g}_0+\check{g}_1+ \check{g}_2$.
The non-perturbative part $\check{g}_0$ can be easily calculated from the standard Eilenberger equation without quantum corrections, supplemented with the normalization condition, $\check{g}_0^2=-\pi^2$~\cite{richardPRB16}.
The correction terms $\check{g}_1$ and $\check{g}_2$ are obtained from an inhomogeneous Eilenberger equation with leading quantum corrections,
\begin{align}
[(\epsilon+e\bm{v}_F\cdot\bm{A})\tau_3-\check{h},\check{g}_{n}]&+ i\boldsymbol{v}_F\cdot\boldsymbol{\nabla}_{\boldsymbol{r}}\check{g}_n =   \nonumber \\
 \frac{i}{2}\{\check{h}\cdot\check{g}_{n-1}\}& - \frac{i}{2}\{\check{g}_{n-1}\cdot\check{h}\}.
\label{eq:ee2}
\end{align}
The thermal conductivity $\kappa = J^z_H/(-\partial _zT)$ is obtained by substituting the solution of $\check{g}=\check{g}_0+\check{g}_1+\check{g}_2+\cdots$ to Eq.~\eqref{eq:hc2}. The temperature gradient along the vortex line is incorporated as the boundary condition of the Keldysh component at $z=\pm\infty$, $g^{K}_n(\infty)=-2\pi(g^{R}_n-g^{A}_n)\tanh[\epsilon/2T(\pm \infty)]$~\cite{SM}, where $g^{R,A}_n$ are calculated in the absence of the temperature gradient.

\begin{figure}
\begin{center}
\includegraphics[width=70mm]{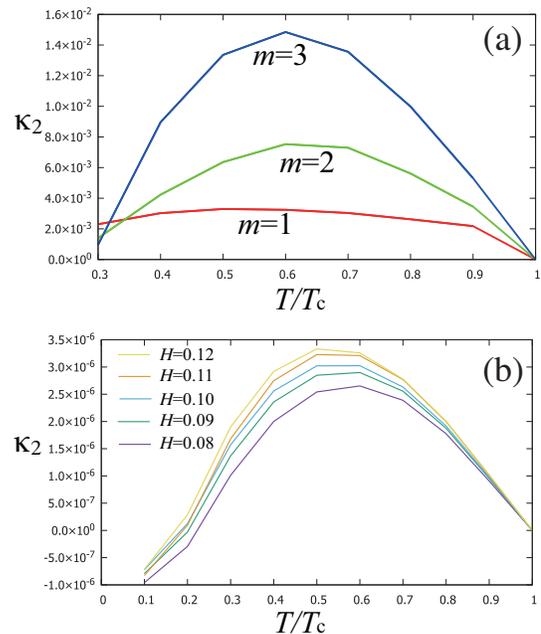}
\end{center}
\caption{(a) $\kappa_2$, versus $T$ in the case of single vortex with vorticity $m=1,2,3$. 
(b) $\kappa_2$ versus $T$ in the case of a vortex lattice for
$H=0.08,~0.09~,~0.10,~0.11,~0.12$ from bottom to top.
}
\label{kappa-1}
\end{figure}

We, first, consider the case of single vortex with vorticity $m$, i.e. 
$\Delta(\bm{r})=\Delta_0(T)[\tanh(r/\xi)]^{|m|}e^{im \phi}$ with $r=\sqrt{x^2+y^2}$.
In this case, we can neglect the vector potential $\bm{A}$ in Eq.~(\ref{eq:ee2}).
Solving Eq.~(\ref{eq:ee2}) numerically for $\check{g}_1$ and $\check{g}_2$, we  found that the contribution from  $\check{g}_1$ 
to the thermal current is negligible. The leading quantum correction associated with the vortex-induced emergent magnetic field
arises from $\check{g}_2$.
The calculated results of this quantum correction term of the thermal conductivity, $\kappa_2$, for vorticity $m=1,2,3$ are shown in FIG.~\ref{kappa-1}(a), where $\kappa _2$ is spatially averaged over the core region within $r\le 5\xi$. In this calculation, the BCS-type temperature-dependence of the gap function is assumed,
the energy unit is scaled by $2\pi T_c$, and
 the parameters are set as, $v_F=20$, $k_F=1$, $\xi=20$, $\Delta_0(0)=1.765T_c$, and $1/\tau=0.002$.
It is noted that $\kappa_2$ increases as the vorticity increases.
Since the emergent magnetic field is proportional to the vorticity, this behavior implies negative magnetoresistivity of thermal currents.
Furthermore, the $T$-dependence of $\kappa_2$ remarkably exhibits upturn increase in the intermediate temperature region, which is indeed in agreement with the prediction from the semiclassical analysis, Eq.~(\ref{eq:tc2}).
However, in contrast to the semiclassical result, which fails down in the low temperature limit,
the $T$-dependence turns to decreasing behaviors in the low temperature region, which 
is consistent with the thermodynamics third law.
Thus, it is concluded that the negative magnetoresistivity of thermal currents is a signature of chiral anomaly of Weyl quasiparticles.
We, here, comment on $T$-dependence of the normal self-energy neglected in our calculations.
If one takes into account the $T$-dependence due to the energy dependence of the density of states, the increase of the thermal conductivity is more magnified in the intermediate $T$-region, because of the longer relaxation time.
Thus, the detection of the chiral anomaly effect becomes more feasible.

We, next, performed the calculation for the case of a vortex lattice. 
For simplicity, a square lattice structure of vortices is assumed \cite{SM,ichiokaPRB02}.
The calculated results of $\kappa_2$ are shown in FIG.~\ref{kappa-1}(b), which is the spatially averaged value over the unit cell.
The qualitative characteristic features are similar to the results for the case with single vortex.
The thermal conductivity increases as a function of a magnetic field, and the $T$-dependence qualitatively coincides
with the Berry phase formula (\ref{eq:tc2}) in the intermediate $T$-region, signifying the chiral anomaly effect.
We also calculated the spatial distribution of thermal currents, and found that 
thermal currents are mainly carried by bulk quasiparticles,
rather than bound states in vortex cores, confirming that
the increase of $\kappa_2$ is due to chiral anomaly of Weyl quasiparticles.
 It is noted that the NTMR in this scenario is free from the issue of current jetting, which disturbs
 the detection of negative magnetoresistivity as a signature of chiral anomaly in the case of Weyl semimetals \cite{armitage}.
 The current jetting is caused by inhomogeneity of current distribution due to the strong Landau quantization.
 Since the wave function in the vortex state is the Bloch function, the current jetting is absent in this case.
We stress that the characteristic temperature dependence found in FIG.\ref{kappa-1} can not be realized for any non-Weyl (non-Dirac) superconductors,
as revealed by numerous previous studies on thermal transport in the vortex state \cite{maki1,maki2,maki3,vek1,vekhter,franz-high,vafek,tesa,takigawa,matsuda,adachi,voron,golubov}.
Thus, the NTMR with the characteristic temperature dependence is a unique feature of Weyl (Dirac) superconductors.

Although the above results establish the NTMR  as a signature of chiral anomaly,
the chiral anomaly contribution shown in FIG.~\ref{kappa-1}(b), which corresponds to the case of high magnetic fields, 
is about 0.1 $\%$ of the total contribution. The calculation for low fields is not 
attainable because of numerical costs.
It is known that for small magnetic fields close to a lower critical field and for $\bm{J}_H \parallel \bm{H}$, the field dependence of the thermal conductivity due to usual pair-breaking is quite small.
Thus, in this case, the experimental detection of the chiral anomaly contribution is still feasible by measuring the field-dependent part of
the thermal conductivity.
A more promising approach for the detection of the chiral anomaly effect is to utilize an emergent chiral magnetic field
induced by lattice strain. 
We consider this scenario in the following.

\begin{figure}
\begin{center}
\includegraphics[width=70mm]{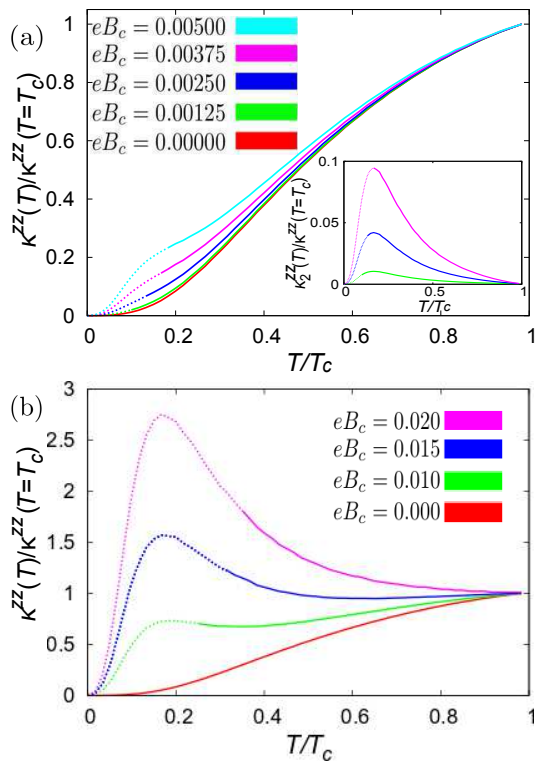}
\end{center}
\caption{(a) $\kappa$ versus $T$ for $e\mathcal{B}_C=0.0, ~0.00125,~0.0025, ~0.00375,~0.005$ from bottom to top. 
For the temperature region $T<  T_L$, in which the quasiclassical approximation fails down, the results are shown in dotted lines.
Inset: $\kappa_2$ versus $T$ for  $e\mathcal{B}_C= 0.00125,~0.0025, ~0.00375,~0.005$ from bottom to top. 
(b) $\kappa$ versus $T$ for $e\mathcal{B}_C=0.0, ~0.01,~0.015,~0.02$ from bottom to top. 
The results for $T<  T_L$ are shown in dotted lines.
}
\label{kappa-2}
\end{figure}

{\it Case of strain-induced chiral magnetic fields.}--- We, now, explore the case that lattice strain induces a chiral magnetic field $\bm{\mathcal{B}}_{\rm C}$ in the 3D chiral $p_x+ip_y$-wave spinless superconductor. 
To simplify the analysis, we introduce the strain-induced chiral vector potential by hand in the mode, though
the realization of the strain-induced magnetic field requires multi-orbital degrees of freedom \cite{PhysRevB.92.165131,PhysRevX.6.041021}.
Since a chiral magnetic field causes neither the Meissner effect nor the vortex state, the pair-breaking effect due to the chiral magnetic field is remarkably weak \cite{SM}. In fact, for the parameters used in our calculations, the superconducting state survives against a chiral magnetic $e\mathcal{B}_C \sim < 0.03$, and thus, we can expect enormous NTMR due to a large 
value of $\mathcal{B}_C$.
The chiral magnetic field in superconductors gives rise to
a pseudo-Lorentz force, which 
is obtained from
the right-hand side of Eq.(\ref{eq:ee1}) \cite{matsushita}.  
For simplicity, we assume a uniform chiral magnetic field parallel to $z$-axis, $\bm{\mathcal{B}}_{\rm C}=(0,0,\mathcal{B}_C)$. Then,
we end up with the Eilenberger equation,
\begin{eqnarray}
[\epsilon \tau_3 - \check{h}, \check{g}]+ i\bm{v}_{\rm F}\cdot{\nabla}_{\bm{r}}\check{g}
+ie\bm{v}_{\rm F}\times\bm{\mathcal{B}}_{\rm C}\cdot\frac{\partial}{\partial\bm{k}_{\parallel}}\check{g} 
 =0.
 \label{eq:ee-cm}
\end{eqnarray}
The last term of (\ref{eq:ee-cm}) is the pseudo-Lorentz force term.
Since this equation is homogeneous, we need an additional normalization condition for $\check{g}$ to solve it, i.e., 
$\check{g}^2=-\pi^2$.
To derive an approximate analytic solution of (\ref{eq:ee-cm}),
we expand $\check{g}$ in terms of $1/(\xi k_F)$ and 
$\bm{\mathcal{B}}_{\rm C}$ up to the second order.
An explicit expression for quantum corrections of $\check{g}$ due to  $\bm{\mathcal{B}}_{\rm C}$ is given in
Supplemental Material \cite{SM}.
Although the superconducting state is robust against large values of $\mathcal{B}_C$, one can not neglect
the Landau quantization of quasiparticles for a sufficiently strong chiral magnetic field, which can not be treated within the quasiclassical approximation.
Thus, the temperature range in which our method is valid is limited to $T>T_L\equiv \sqrt{2e\mathcal{B}_C}\Delta/k_F$, for which the Landau levels
are smeared by temperature broadening effect.
We calculate a thermal current from Eq.(\ref{eq:hc2}) up to linear order in $\nabla T$ \cite{graf,voron}.
Numerical results of the thermal conductivity $\kappa=\kappa_0+\kappa_2$ with $\kappa_0$ 
the non-perturbed zero-field part and $\kappa_2$
the field-dependent quantum correction,
are shown in FIG.\ref{kappa-2}.
In this calculation, the BCS-type $T$-dependence of the gap function, and the same parameters as those in the case with
vortex-induced magnetic fields are used.

As seen in FIG.\ref{kappa-2}, the thermal conductivity increases, as $\mathcal{B}_C$ increases, signfying NTMR. Furthermore, for $e\mathcal{B}_C >  \sim0.01$, 
the quantum correction part dominates over, and hence,
the total thermal conductivity exhibits a remarkable increase, as temperature is lowered in the intermediate temperature region, which
is a characteristic feature of chiral anomaly contributions. 
The positions of the peaks of $\kappa$ for different values of $\mathcal{B}_C$ shown in FIG.\ref{kappa-2} (b) are roughly 
$T_c \times \Delta /E_F$, and thus independent of $\mathcal{B}_C$.
It is noted that the prominent increase of the thermal conductivity appears even for temperatures much above $T_L$ for sufficiently 
large $e\mathcal{B}_C$, implying that
 the increasing behavior of the thermal conductivity is not an artifact of the quasiclassical approximation.
For putative Weyl superconductors of uranium-based systems with lattice constants $4\sim 9$ \AA, $\mathcal{B}_C \approx  2 \sim 5$ Tesla (T) can be realized
by torsional distortion around the $c$-axis by $2\pi$ per $\sim 1$ $\mu$m.  
On the other hand, for a lattice constant $\sim 4$ \AA, $e\mathcal{B}_C=0.00125$ in FIG.\ref{kappa-2} corresponds to $\mathcal{B}_C\sim 5$ T.
In such cases, the magnitude of the chiral anomaly part of the thermal conductivity is more than 10 $\%$ of the 
total thermal conductivity, 
and thus,
it is feasible to detect the characteristic $T$-dependence of $\kappa_2$ experimentally
by extracting $\mathcal{B}_C$-dependent part 
of the thermal conductivity.
We also note that current jetting issue \cite{armitage} can be avoided in this case, because the results in FIG. \ref{kappa-2} shows that the characteristic signature of chiral anomaly, i.e. the upturn increase of the thermal conductivity in the intermediate temperature region, appears even for sufficiently small chiral magnetic fields which do not cause the inhomogeneous current distribution due to the strong Landau quantization.

{\it Conclusion.}--- We have investigated thermal transport in Weyl superconductors with
emergent (chiral) magnetic fields. It is established that NTMR as a signature of chiral anomaly of Weyl quasiparticles can be realized, and its experimental detection is feasible.

This work was supported by the Grant-in-Aids for Scientific
Research from MEXT of Japan [Grants No.~JP17K05517, No.~25220711, and No.~JP16K05448] and KAKENHI on Innovative Areas ``Topological Materials Science'' [No.~JP15H05852, No.~JP15H05855] and "J-Physics" [No.~JP18H04318]. 

\bibliography{NTM}
\bibliographystyle{apsrev}


\newpage

\begin{center}
{\LARGE Supplemental Material}
\end{center}

\appendix

\section{Spatial distortion, vielbein, and torsion field}
Effects of vortex textures and lattice strain which generate emergent chiral magnetic fields in Weyl/Dirac fermion systems can be treated by using a vielbein field $e^{\mu}_a$ with $\mu=1,2,3$ space indices for the laboratory frame, and $a=\bar{1},\bar{2},\bar{3}$ indices for a local orthogonal frame \cite{PhysRevB.92.165131,T.Hughes2011,T.Hughes2013,Parrikar2014,Gromov2014,PhysRevLett.115.177202,PhysRevLett.116.166601,PhysRevB.94.241405,PhysRevX.6.041046,PhysRevX.6.041021,PhysRevB.96.081110,PhysRevB.95.041201,PhysRevB.96.224518}. The vielbein relates the laboratory frame with spatial distortion $x^{\mu}$ with the local orthogonal frame $X^a$
as $d x^{\mu}=e^{\mu}_a dX^a$, or $d X^{a}=e^{a}_{\mu} dx^{\mu}$ where $e^{a}_{\mu}$ is the inverse of $e^{\mu}_a$. If the spatial distortion is sufficiently small, the vielbein is expanded as $e^{a}_{\mu}\approx \delta^{a}_{\mu}-\partial_{\mu}u^{a}$, where $u^{a}$ is the displacement vector parameterizing the spatial distortion.
Then, a momentum operator $k_{\mu}$ is replaced by
\begin{eqnarray}
e^{a}_{\mu}k_{a}\approx k_{\mu}-\partial_{\mu} u^{a} k_{a}.
\label{eq:sup1}
\end{eqnarray} 
The second term of the right-hand side of Eq.(\ref{eq:sup1}) plays a role similar to a U(1) gauge field.
In fact, since it depends on momentum $k_a$, i.e. the position of Weyl points in the momentum space, it can be regarded as
a chiral vector potential, the coupling charge of which depends on chirality of Weyl fermions. 
When the rotation of this emergent chiral vector potential is nonzero, it gives a chiral magnetic field.
More precisely, the chiral magnetic field is expressed as,
\begin{eqnarray}
\mathcal{B}^{\mu}=\frac{\epsilon^{\mu\nu\lambda}}{2}T^a_{\nu\lambda}k_a,
\label{eq:sup2}
\end{eqnarray}
where a torsion field $T^a_{\mu\nu}$ is defined as,
\begin{eqnarray}
T^a_{\mu\nu}=\partial_{\mu}e^a_{\nu}-\partial_{\nu}e^a_{\mu}.
\label{eq:sup3}
\end{eqnarray}

\section{Semiclassical equation of motion with Berry curvatures}
Semiclassical equation of motion with the Berry curvature in the momentum space 
characterizing Weyl quasiparticles, and a torsion-induced emergent magnetic field
can be obtained from the path-integral formalism  by applying adiabatic approximation,
\begin{eqnarray}
\frac{d \bm{r}}{dt}&=&
\frac{\partial E_{\bm{k}s}}{\partial \bm{k}}
+\frac{\partial U(\bm{r})}{\partial \bm{r}}\times \bm{\Omega}_{\bm{k}\bm{k}s} -(\frac{\partial E_{\bm{k}s}}{\partial \bm{k}}\cdot\bm{\Omega}_{\bm{k}\bm{k}s})\bm{\mathcal{B}}  \nonumber \\
&&-\hat{\Omega}_{\bm{k}\bm{r}s}\cdot\frac{d\bm{r}}{dt}
+(\bm{\Omega}_{\bm{k}\bm{k}s} \cdot \bm{\mathcal{B}})\frac{d \bm{r}}{dt},
\label{eq:dr}
\end{eqnarray}
\begin{eqnarray}
\frac{d\bm{k}}{dt}&=&-\frac{\partial U(\bm{r})}{\partial \bm{r}}
+\frac{d\bm{r}}{dt}\times \mathcal{\bm{B}}
+(\frac{\partial U(\bm{r})}{\partial \bm{r}}\cdot \bm{\mathcal{B}})\bm{\Omega}_{\bm{k}\bm{k}s} \nonumber \\
&+&\hat{\Omega}_{\bm{r}\bm{k}s}\cdot\frac{d\bm{k}}{dt} 
+(\bm{\Omega}_{\bm{k}\bm{k}s} \cdot \bm{\mathcal{B}})\frac{d \bm{k}}{dt},
\label{eq:dp}
\end{eqnarray}
where $\bm{\mathcal{B}}$ is an emergent (chiral) magnetic field, and,
 \begin{eqnarray}
\bm{\Omega}_{\bm{k}\bm{k}s}=i\langle \nabla_{\bm{k}} u_{s}| \times | \nabla_{\bm{k}} u_{s}\rangle,
\end{eqnarray}
\begin{eqnarray}
(\hat{\Omega}_{\bm{k}\bm{r}s})_{\alpha\beta}=i(\langle \partial_{k_{\alpha}} u_{s}| \partial_{r_{\beta}} u_{s} \rangle -
\langle \partial_{r_{\beta}} u_{s}| \partial_{k_{\alpha}} u_{s} \rangle),
\end{eqnarray}
\begin{eqnarray}
\hat{\Omega}_{\bm{r}\bm{k}s}=-(\hat{\Omega}_{\bm{k}\bm{r}s})^t,
\end{eqnarray}
with $|u_{s}\rangle$ the eigen function of the lower band of Weyl fermions with the monopole charge $s$,
and $E_{\bm{k}s}=-\varepsilon_{\bm{k}s}$.
Here, $U(\bm{r})$ is a fictitious gravitational potential which induces thermal current flow.
In the derivation of the above equation of motions, we retained the terms that are at most the first order in the Berry curvatures.
This approximation is sufficient for our purpose.
The spatial gradient of $U(\bm{r})$ is associated 
with temperature gradient $-\nabla U \rightarrow -E_{\bm{k}s}\nabla T /T$.
Using the above semiclassical equation of motion and the Boltzmann equation, we obtain
the thermal current given by Eq.(4) in the main text. 

\section{Keldysh formalism of the Eilenberger equation}
We, here, present the detail of the Keldysh formalism of the Eilenberger equation.
The quasiclassical Green function is defined by,
\begin{eqnarray}
\check{g}(\hat{\boldsymbol{k}},\boldsymbol{r},\epsilon)
=
	\begin{pmatrix}
	\hat{g}^R & \hat{g}^K \\
	0 &\hat{g}^A \\
	\end{pmatrix}
,
\end{eqnarray}
\begin{eqnarray}
\hat{g}^{R,A}=
\begin{pmatrix}
	g^{R,A} & f^{R,A}\\
	\tilde{f}^{R,A} & \tilde{g}^{R,A}\\
\end{pmatrix}
, \quad
\hat{g}^{K}=
\left(
\begin{array}{cc}
	g^{K} & f^{K}\\
	-\tilde{f}^{K} & -\tilde{g}^{K}\\
\end{array}
\right),
\end{eqnarray}
and,
\begin{eqnarray}
\tilde{g}^X(\hat{\bm{k}},\bm{r},\epsilon)=g^{X \dagger}(-\hat{\bm{k}},\bm{r},-\epsilon),
\end{eqnarray}
with $X=R,A,K$.
Taking into account quantum corrections to quasiclassical approximation up to leading order,
we have the Eilenberger equation satisfied by the quasiclassical Green function,
\begin{align}
\begin{split}
[(\epsilon+e\bm{v}_{F}\cdot\bm{A})\tau_3-\check{h},\check{g}]+i\boldsymbol{v}_F\cdot\boldsymbol{\nabla}_{\boldsymbol{r}}\check{g}=\frac{i}{2}\{\check{h}\cdot\check{g}\}-\frac{i}{2}\{\check{g}\cdot\check{h}\}, 
\end{split}
\label{eq:a-ei}
\end{align}
where
\begin{eqnarray}
\{\check{a}\cdot\check{b}\}&=\boldsymbol{\nabla}_{\boldsymbol{r}}\check{a}\cdot\boldsymbol{\nabla}_{\boldsymbol{k}}\check{b}-\boldsymbol{\nabla}_{\boldsymbol{k}}\check{a}\cdot\boldsymbol{\nabla}_{\boldsymbol{r}}\check{b},
\end{eqnarray}
and $\check{h}=\check{\Delta}+\check{\sigma}_{\rm imp}$ with,
\begin{eqnarray}
\check{\Delta}
=
	\begin{pmatrix}
	\hat{\Delta}^R & \hat{\Delta}^K \\
	0 &\hat{\Delta}^A \\
	\end{pmatrix}
,\quad
\check{\sigma}_{\rm imp}
=
	\begin{pmatrix}
	\hat{\sigma}_{\rm imp}^R & \hat{\sigma}_{\rm imp}^K \\
	0 &\hat{\sigma}_{\rm imp}^A \\
	\end{pmatrix}
,
\end{eqnarray}
and,
\begin{eqnarray}
\hat{\Delta}^{R,A}
=
	\begin{pmatrix}
	0 & \Delta^{R,A} \\
	\tilde{\Delta}^{R,A} &0 \\
	\end{pmatrix}
,\quad
\hat{\Delta}^{K}
=
	\begin{pmatrix}
	0 & \Delta^{K} \\
	-\tilde{\Delta}^{K} &0 \\
	\end{pmatrix}
,	
\end{eqnarray}
\begin{eqnarray}
\hat{\sigma}_{\rm imp}^{R,A}
=
	\begin{pmatrix}
	\sigma_{\rm imp}^{R,A} & 0 \\
	0 &\tilde{\sigma}_{\rm imp}^{R,A} \\
	\end{pmatrix}
, \quad
\hat{\sigma}_{\rm imp}^{K}
=
	\begin{pmatrix}
	\sigma_{\rm imp}^{K} & 0 \\
	0 & -\tilde{\sigma}_{\rm imp}^{K} \\
	\end{pmatrix}
	.
\end{eqnarray}
Since the energy dependence of the gap function is neglected, we have,
\begin{eqnarray}
\Delta^R=\Delta^A=\Delta (\bm{r}) (k_x-i k_y)/k_F,
\end{eqnarray}
within quasiclassical approximation.
The retarded and advanced self-energy due to impurity scattering within the Born approximation is given by,
\begin{eqnarray}
\sigma^{R,A}_{\rm imp}=\frac{1}{2\pi \tau}\langle g^{R,A}\rangle ,
\label{eq:sigma}
\end{eqnarray}
where $\tau$ is the relaxation time of quasiparticles, and $\langle ... \rangle$ is the average over the Fermi surface.
In the calculations of the main text, we consider leading corrections of thermal conductivity due to emergent (chiral) magnetic fields.
Thus, we replace $g^{R,A}$ in Eq.(\ref{eq:sigma}) with those without quantum corrections, i.e.,  the solutions of (\ref{eq:a-ei}) with the
right-hand side equal to zero.
The deviation from equilibrium due to temperature gradient mainly affects the distribution function of quasiparticles for thermal transport. Thus, we neglect the deviation from the equilibrium values of the gap amplitude,
putting $\Delta^K=0$.
Because of the same reasons, the Keldysh part of the self-energy is replaced by the equilibrium one
without quantum corrections,
\begin{eqnarray}
\label{eq: equilibrium Green function}
\sigma^K_{\rm imp}=(\sigma^R_{\rm imp}-\sigma^A_{\rm imp})\tanh\left(\frac{\epsilon}{2T}\right).
\end{eqnarray}

\section{Vortex lattice structure}

\begin{figure}[t]
\includegraphics[width=80mm]{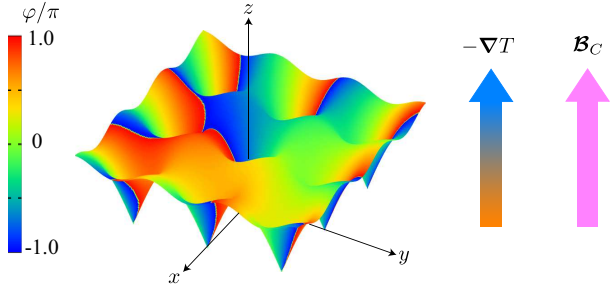}
\caption{Spatial profile of the gap amplitude, $|\Delta({\bm r})|$, and the phase, $\varphi({\bm r})$, (color map) in a square vortex lattice. The temperature gradient is applied along the vortex line and the torsional magnetic field due to the vortex lattice emerges along the same direction.}
\label{fig:lattice}
\end{figure}

In the main text, we consider the thermal transport in the presence of a square array of vortices. Following Ref.~\cite{ichiokaPRB02}, we introduce the unit vectors of a square vortex lattice as ${\bm a}_1=(a_x,0)$ and ${\bm a}_2=(0,a_y)$, where we use the relation $Ha_xa_y=\Phi _0$ and $\Phi _0$ is the flux quantum. The spatial coordinate in a unit cell is parameterized with $u\in[-1/2,1/2]$ and $v\in[-1/2,1/2]$ as ${\bm r}=u{\bm a}_1+v{\bm a}_2$ and we impose the periodic boundary condition, 
\beq
\Delta ({\bm r}+{\bm R}) = \Delta({\bm r})e^{i\chi ({\bm r})},
\eeq
and ${\bm A}({\bm r}+{\bm R})={\bm A}({\bm r})$ with the lattice translation vector ${\bm R}=m{\bm a}_1+n{\bm a}_2$ ($m,n\in \mathbb{Z}$). In the symmetric gauge, ${\bm A}({\bm r})=\frac{1}{2}{\bm H}\times {\bm r}+{\bm a}({\bm r})$ and the phase factor is given by
\beq
\chi ({\bm r}) =  - \frac{\pi}{\Phi _0}\left( {\bm H}\times {\bm R}\right)\cdot\left( {\bm r}+2{\bm r}_0\right)
-\pi mn .
\eeq
In each unit cell, we consider a singly quantized vortex centered at $u=v=0$, corresponding to ${\bm r}_0=({\bm a}_1+{\bm a}_2)/2$. For simplicity, we neglect the spatial inhomogeneity of the internal field, i.e., ${\bm a}={\bm 0}$. Figure~\ref{fig:lattice} shows the spatial profiles of the gap amplitude and phase in a square vortex lattice. The phase gradient generates the emergent magnetic field $\bm{\mathcal{B}}$ along the vortex line which is parallel to the temperature gradient. 

\section{Thermal conductivity in the case of a uniform chiral magnetic field}

In this section, we present the derivation of approximated solutions of the Eilenberger equation with a pseudo-Lorentz force term, Eq.(11), in the main text, and 
the formula of the thermal conductivity, which are used in the calculation of the main text.

We derive the quantum correction of the quasiclassical Green function using perturbative calculation with respect to a temperature gradient and a chiral magnetic field. The quasiclassical Green function and the self-energy can be written as,
\begin{gather}
\check{g}=\check{g}_{{\rm eq}}+\delta\check{g},\\
\check{\sigma}_{{\rm imp}}=\check{\sigma}_{{\rm imp\;eq}}+\delta\check{\sigma}_{{\rm imp}}.
\end{gather}
where $\check{g}_{{\rm eq}}$ and $\check{\sigma}_{{\rm imp\;eq}}$ are the quasiclassical Green function and the self-energy in local equilibrium, and $\delta \check{g}$ and $\delta\check{\sigma}_{{\rm imp}}$ are the deviation from equilibrium.
As mentioned above, we neglect $\delta\check{\sigma}_{{\rm imp}}$  in the following.
Up to the first order in the temperature gradient, the deviation from equilibrium of the Keldysh Green function is given by,
\beq
\label{Elia}
\delta g^K=(\delta g^R-\delta g^A)\tanh \left(\frac{\e}{2T} \right)+\delta g^a,
\eeq
where $\delta g^a$ is the anomalous part which describes non-equiliburium effects. We parametrize the anomalous function as,
\beq
\delta g^a \equiv \bpm
g^a+g'^{a}&&f^a\\
-f^{a\dag}&&g^a-g'^{a}
\epm.
\label{eq:anoma}
\eeq
Note that only $g_a$ in Eq.(\ref{eq:anoma}) contributes to the thermal conductivity, as will be seen below.
Since we consider anisotropic pairing states, it is legitimate to assume that the off-diagonal part of the self-energy due to impurity scattering is negligibly small within the Born approximation. Using a manner similar to developed in Ref. \cite{graf,voron}, 
up to the first order in the temperature gradient, we obtain,
\beq
\label{ga}
g^a= -i{\bm v}_F\cdot \nabla T \left( \frac{\p}{\p T}\tanh \frac{\e}{2T}\right) \frac{g_{{\rm eq}}^R-g_{{\rm eq}}^A}{\s_{{\rm imp\;eq}}^R-\s_{{\rm imp\;eq}}^A }.
\eeq
Here, we have neglected the off-diagonal components of the self-energy and used the relation of the equilibrium Green function: $f^R_{{\rm eq}}=-f^{\dag A}_{{\rm eq}}$.  

We, now, derive an analytical expression for thermal conductivity with the second-order quantum correction due to a uniform chiral magnetic field $\bm{\mathcal{B}}_{\rm C}=(0,0,\mathcal{B}_C)$.
The definition of the thermal conductivity $\kappa$ is given as,
\begin{align}
\label{thermoco}
{\bm J}_{\rm H}({\bm r}) &\equiv -\kappa\nabla T \nn \\
&=N_F\int^\infty_{-\infty}\frac{d\e}{4\pi i}\int d\hat{{\bm k}}\e {\bm v}_{ F} \frac{1}{2}{\rm Tr}\delta\underline{g}^K(\hat{{\bm k}},{\bm r},\e).
\end{align}
We can obtain the trace of the Keldysh Green function and  the analytical expression of the thermal conductivity Eq. (\ref{thermoco}) from the Eq. (\ref{ga}). 
Then, the thermal conductivity is given by,
\begin{align}
\label{eq:kappa-chiral}
\kappa_{ij}(T)=\frac{N_F}{2T^2}\int\frac{d\Omega_k}{4\pi}\frac{d\epsilon}{4\pi}\frac{\epsilon^2 v_{Fi}v_{Fj}}{\sigma^R_{\rm imp\;eq}-\sigma^A_{\rm imp\;eq}}\frac{g_{{\rm eq}}^R-g_{{\rm eq}}^A}{\cosh^2\frac{\epsilon}{2T}},
\end{align}
where the subscript $i,j=x,y,z$.
The effects of the chiral magnetic field are incorporated in the retarded and advanced Green function. Performing analytical continuation of the Matsubara Green's functions which are expanded up to the second-order in the chiral magnetic field, 
we obtain the retarded Green function,
\begin{align}
g^R_{{\rm eq}}(\e)&=-i\pi\frac{( \e-\s_{\rm imp\;eq}^R )}{Z^R(\e)}e^{-\frac{i(\theta_1^R+\theta_2^R)}{2}}\nn\\
&-i\pi\frac{ev_F{\mathcal{B}_{\rm C}}|\D|^2 \sin^2 \theta_k}{2k_F[Z^{R}(\e)]^3}e^{-\frac{3i(\theta_1^R+\theta_2^R)}{2}}\nn\\
&+i\pi\frac{e^2v_F^2{\mathcal{B}_{\rm C}^2}|\D|^2 \sin^2 \theta_k(\e-\s_{\rm imp\;eq}^R)}{8k_F^2[Z^{R}(\e)]^5}e^{-\frac{5i(\theta_1^R+\theta_2^R)}{2}},
\label{retarted function}
\end{align}
where the coefficient and  phases are defined as : $Z^R(\e)\equiv |(\e-\s_{\rm imp\;eq}^R )^2-|\D\sin \theta_k|^2|^\frac{1}{2}$ and $\theta^R_1(\e)={\rm arg}(\e-\s_{\rm imp\;eq}^R+|\D \sin \theta_k|),\; \theta^R_2(\e)={\rm arg}(\e-\s_{\rm imp\;eq}^R-|\D \sin \theta_k|)$. The advanced Green function can be obtained from the retarded Green function : $g^A_{{\rm eq}}(\e)=[g^R_{{\rm eq}}(\e)]^{*}$.
The numerical results of the thermal conductivity shown in FIG.2 in the main text are obtained by using Eq. (\ref{eq:kappa-chiral}) and Eq. (\ref{retarted function}).

\section{Pair breaking effect due to chiral magnetic field}
A pair breaking effect due to a chiral magnetic field has been discussed before on the basis of a microscopically derived Ginzbur-Landau equation \cite{matsushita}. 
The Ginzburg-Landau equation for a chiral $p_x+ip_y$-wave superconductor with a chiral magnetic field is given by \cite{matsushita},
\begin{widetext}
\begin{eqnarray}
\label{GL}
(1-\frac{T}{T_c})\Delta(\bm{r})-\frac{7\zeta(3)}{10\pi^2 T_c^2}|\Delta(\mbox{\bm $x$})|^2 \Delta(\bm{r})+\frac{7\zeta(3)v_F^2}{40\pi^2 T_c^2}(\partial_x^2+\partial_y^2+\frac{1}{2}\partial_z^2)\Delta(\bm{r})+\frac{7\zeta(3)v_F^2}{64\pi^2 T_c^2}\left(3i\frac{e\mathcal{B}_{\rm C}}{k_F}\partial_z-4\left( \frac{e^2\mathcal{B}_{\rm C}^2}{k_F} \right)^2\right)\Delta(\bm{r})=0,\nn\\
\end{eqnarray}
\end{widetext}
where $\partial_i \equiv \frac{\partial}{\partial r_i}-2ieA_i(x) \;(i=x,y,z)$ is the gauge invariant differential operator. The fourth term stabilizes the Fulde-Ferrell state with $\D(\bm{r})=|\D|e^{iQz}$ \cite{matsushita}. To discuss the pair breaking effect due to a chiral magnetic field, we determine the center of mass momenta $Q$ and the critical temperature from the stationary condition of the Ginzburg-Landau free energy. Under this condition, we obtain $Q=-\frac{15}{4}\frac{e\mathcal{B}_{\rm C}}{k_F}$, and the critical temperature given by,
\begin{eqnarray}
\label{delTc}
\frac{\delta T_c}{T_c}=-0.166\left( \frac{ev_F\mathcal{B}_{\rm C}}{k_FT_c} \right)^2.
\end{eqnarray}
Here, $T_c$ is a critical temperature in the case without chiral magnetic fields, and $\delta T_c$ is the variation of the critical temperature due to the chiral magnetic field. 
Eq. (\ref{delTc}) implies that the critical temperature is decrease by the chiral magnetic field. In the case of $\mathcal{B}_{\rm C}=0.02$, which corresponds to 16 times larger than the upper critical magnetic field $H_{c2},\; \delta T_c/T_c \sim -1$. 
Thus, $\mathcal{B}_{\rm C}\approx 0.02$ gives a rough estimate of the critical chiral magnetic field. 
Therefore, the superconducting state survives even for a strong chiral magnetic field, the magnitude of which is 10 times lager
than the upper critical field. 
We can conclude that the pair breaking effect due to the chiral magnetic field is very weak compared to that due to a usual magnetic field.

\section{Relation and comparison with previous studies}
There have been several related works on chiral anomaly phenomena in Weyl systems. We, here, discuss the relation and comparison between our approaches and previous ones. The chiral anomaly originates from the violation of the conservation law of the axial current in the presence of both an electric field and a magnetic field, or analogous external fields caused by geometrical responses \cite{PhysRevB.92.165131,T.Hughes2011,T.Hughes2013,Parrikar2014,Chandia,Shitade2014,Gromov2014,PhysRevLett.115.177202,PhysRevLett.116.166601,PhysRevB.94.241405,PhysRevX.6.041046,PhysRevX.6.041021,PhysRevB.96.081110,PhysRevB.95.041201,PhysRevB.96.224518,Lucas9463}. In the case of Weyl superconductors, chiral anomaly induced by geometrical responses is relevant, because Weyl-Bogoliubov quasiparticles do not directly couple to usual electromagnetic fields.
In particular, emergent chiral magnetic fields generated by vortex textures and lattice strain are described  in terms of torsion fields
Eq.(\ref{eq:sup3}). It is noted that the so-called "gravitomagnetic field" considered in \cite{PhysRevLett.108.026802,Gromov2014} corresponds to the emergent chiral magnetic field given by Eq.(\ref{eq:sup2}) in Weyl superconductors.
 Then, the torsion field modifies the conservation law of the axial current, as clearly shown by You et al. \cite{PhysRevB.94.085102} (Eq.(30) of this reference). 
It is noted that the chiral anomaly due to the torsion field implies the existence of the axion-type effective action expressed in terms of torsion fields, i.e. the Nieh-Yan term which was discussed by Chandia and Zanelli \cite{Chandia}.
 Combining this modified conservation law with a transport theory such as a hydrodynamis argument given by Lucas et al. \cite{Lucas9463}, one may obtain the same results as our findings. The calculation based on this approach is straightforward when the chiral magnetic field is uniform. However, in the case of texture-induced emergent magnetic field, which is considered in the second part of the main text, the emergent field is generally inhomogeneous, and thus the calculation would be rather involved. In such cases, our approach based on the Eilenberger equation is  advantageous for the calculation of transport properties.

\section{Thermal conductivity for non-Weyl (non-Dirac) superconductors}
We, here, compare the thermal transport of usual superconductors in which there is no Weyl (or Dirac) quasiparticles with
that of Weyl (Dirac) superconductors. The purpose of this section is to clarify that the signature of thermal conductivity shown in FIG.1 in the main text can not be realized for non-Weyl (non-Dirac) superconductors, and to confirm that the signature is a unique feature of Weyl (Dirac) superconductors. 
According to numerous previous studies \cite{maki1,maki2,maki3,vek1,vekhter,franz-high,vafek,tesa,takigawa,matsuda,adachi,voron,golubov}, in the vortex state of non-Weyl (non-Dirac) superconductors, there are three mechanisms which increase the thermal current as a magnetic field increases: (i) the suppression of the superconducting gap due to pair-breaking effects caused by a magnetic field, (ii) the increase of the density of state of quasiparticle bound states in vortex cores, (iii) the increase of the density of states due to the Doppler shift  (Volovik effect). 
The mechanism (i) is not included in our calculation, and does not explain our findings. For the mechanism (ii), 
we have examined that, in our calculations, the thermal current is dominated by contributions from bulk regions rather than the bound states in vortex cores. 
Thus, the mechanism (ii) does not explain the NTMR found in our calculations for Weyl superconductors. The mechanism (iii) is possible only when an applied magnetic field is perpendicular to the direction along which nodes of the gap open. Thus, this mechanism is not applicable to our setup where the node direction (parallel to $k_z$) is parallel to the magnetic field.
Furthermore, It is well known that all of the above three mechanisms do not give rise to the characteristic temperature dependence of the thermal conductivity found in our paper as a signal of the chiral anomaly contributions, i.e. the increasing behavior in the intermediate temperature region.  Thus, we can definetely conclude that the negative thermal magnetoresistivity with the characteristic temperature dependence is absent for non-Weyl (non-Dirac) superconductors, and that the signal we found in the thermal transport is a unique feature of Weyl (Dirac) superconductors.

\end{document}